\pdfoutput=1
\documentclass[aps, reprint, prd, superscriptaddress, nofootinbib, showpacs, floatgx]{revtex4-1}
\usepackage{amsmath, latexsym, amssymb, hyperref, graphicx, color, multirow}
\usepackage{mathtools,slashed}
\usepackage[table]{xcolor}
\usepackage{tabularx}
\usepackage[T1]{fontenc}
\usepackage[capitalize]{cleveref}
\usepackage{amssymb}
\usepackage{graphicx}
\usepackage{url}
\usepackage{enumitem}
\usepackage{hyperref}
\definecolor{nicered}{rgb}{.7,.1,.1}
\definecolor{nicegreen}{rgb}{.1,.5,.1}
\definecolor{darkblue}{rgb}{0,0,.5}
\hypersetup{colorlinks, citecolor=nicegreen,linkcolor=nicered, urlcolor=darkblue}

\begin{document}


\title{Probing the Dirac  or Majorana nature of the Heavy Neutrinos in pure leptonic decays at the LHC}

\author{Carolina Arbela\'ez}
\email{carolina.arbelaez@usm.cl}
\affiliation{Universidad T\'ecnica Federico Santa Mar\'ia,\\
	Centro-Cient\'ifico-Tecnol\'ogico de Valpara\'iso, Valpara\'iso, Chile}
    \author{Claudio Dib}
    \email{claudio.dib@usm.cl}
\affiliation{Universidad T\'ecnica Federico Santa Mar\'ia,\\
	Centro-Cient\'ifico-Tecnol\'ogico de Valpara\'iso, Valpara\'iso, Chile}
    \author{Iv\'an Schmidt}
       \email{ivan.schmidt@usm.cl}
\affiliation{Universidad T\'ecnica Federico Santa Mar\'ia,\\
	Centro-Cient\'ifico-Tecnol\'ogico de Valpara\'iso, Valpara\'iso, Chile}
\author{Juan Carlos Vasquez}
\email{juan.vasquezcar@usm.cl}
\affiliation{Universidad T\'ecnica Federico Santa Mar\'ia,\\
	Centro-Cient\'ifico-Tecnol\'ogico de Valpara\'iso, Valpara\'iso, Chile}

\begin{abstract}
\noindent
We propose a 
strategy for distinguishing the Dirac / Majorana character of heavy neutrinos  with masses below the $W$ boson mass, using purely leptonic decays at the LHC. The strategy makes use of a forward-backward asymmetry of the opposite charge lepton in the $W^{+}\rightarrow l^{+}l^{+}l^{'-}\nu$ decay. In order to check the experimental feasibility  of the model, we show, through a numerical analysis, that in the decay
$W^{+}\rightarrow e^{+}e^{+}\mu^{-}\nu$ the two positrons in the final state can be distinguished for different ranges of the heavy neutrino masses.
Finally, we estimated the number of events of $W^{+}\rightarrow e^{+}e^{+}\mu^{-}\nu$ for a Dirac and Majorana $N$ neutrino. For an integrated luminosity of 120 fb$^{-1}$ at LHC RUN II, signals can be found if heavy-to-light neutrino mixings are $ |U_{N \mu}|^2,|U_{N e}|^2 \gtrsim 10^{-6} $.

\end{abstract}

\maketitle

\section{Introduction}
\label{sec:intro}

Experiments in the last decades have confirmed that at least two of the three known neutrinos must have nonzero masses, and that all three have non-trivial mixings with respect to electroweak flavors. Ultimate confirmation of neutrino masses came from the results of atmospheric, solar and reactor neutrino experiments \cite{PhysRevLett.81.1562,Ahmad:2002jz,Eguchi:2002dm,Forero:2014bxa}.
%
This evidence, compounded by the fact that the
neutrino  masses happened to be tiny with respect to the other Standard Model (SM) fermions, is  currently  an  outstanding  theoretical path for physics that goes
beyond  the  SM.   Most of the explanations  of the neutrino mass smallness are  based  on  the  existence  of  extra  heavy neutral fermions, which could be Dirac or Majorana \cite{Majorana2008}. In fact, the most widely accepted mechanism to generate small neutrino masses, is the seesaw mechanism \cite{MINKOWSKI1977421,Mohapatra:1979ia,Glashow:1979nm,GellMann:1980vs,Yanagida:1979as}, which involves extra heavy sterile neutrinos (henceforth denoted generically by $N$). Moreover, in most of these scenarios the neutrinos are Majorana instead of Dirac fermions, and this discrimination about their nature is a  crucial and challenging  piece of information  that  experiments  must elucidate.

Direct collider searches for sterile heavy neutrinos may provide
a simultaneous probe of both their Dirac or Majorana nature, as well as their mixing with the active neutrinos. At hadron colliders, a strong signal of heavy Majorana neutrinos has been shown to exist and it is the same-sign dilepton final states, with two jets and no missing transverse energy: $pp\rightarrow W\rightarrow N e^{\pm}\rightarrow e^{\pm}e^{\pm}jj$. This was originally proposed in \cite{Keung:1983uu} and further studied in \cite{Ferrari:2000sp,Kovalenko:2009td,Han:2012vk,Das:2015toa,Das:2016hof,Das:2017gke,Das:2017pvt,Das:2017zjc,Das:2017hmg}. Using this channel, both CMS and ATLAS experiments at the LHC have set direct limits on the light-heavy neutrino mixing \cite{Aad:2015xaa, Khachatryan:2015gha, Atre:2009rg, Helo:2013esa}, for masses of $m_{N}\sim 100 - 500$ GeV.

For neutrino masses in the region $m_{N}< m_{W}$, the produced jets in the final state may not pass the cuts required to reduce the backgrounds, so the purely leptonic channels,  such as $pp \rightarrow  e^{\pm}e^{\pm}\mu^{\mp}\nu$, in which $\nu$ could be either a neutrino or a anti-neutrino, may be more favorable \cite{Izaguirre:2015pga}. Here we will focus on the signal $W^{\pm}\rightarrow e^{\pm}e^{\pm}\mu^{\mp}\nu^{-}$, which would indicate the existence of a sterile neutrino through the subprocess $W^{\pm}\rightarrow e^{\pm}N$, followed by $N\rightarrow e^{\pm}\mu^{\mp}\nu$.
 Notice that in the leptonic channel it is not clear a priori how to distinguish Dirac neutrinos  from Majorana ones, since the final neutrino gets undetected in both cases (i.e, the observed final states are $e^{\pm}e^{\pm}\mu^{\mp}$ or $\mu^{\pm}\mu^{\pm}e^{\mp}$ plus missing energy ).  Hence, an obvious question is whether a Dirac or Majorana $N$ can be distinguished at the LHC in these pure leptonic modes.   

This has been answered in Refs. \cite{Dib:2015oka,Dib:2016wge,Dib:2017iva,Dib:2017vux}, where it was shown that the Dirac and  Majorana cases could in fact be distinguished, depending of the relative size of the mixing  between the $e$ or $\mu$ flavors  with the heavy neutrinos. In particular, they have have addressed the issue of Dirac/Majorana discrimination using the muon spectrum and a  multi-variable analysis. 

Here we present a simpler approach that makes use  of the forward-backward asymmetry proposed in \citep{Tait:2000sh,Atre:2009rg}.
Complementary to the strategy presented in \cite{Dib:2016wge,Dib:2017iva,Dib:2017vux}, we find that a suitable quantity to discriminate between Dirac and Majorana in the purely leptonic channels is the forward-backward asymmetry of the unlike-charged lepton~\cite{Han:2012vk}. This strategy works even when the mixing of the heavy neutrinos with the charged leptons are equal or of the same order of magnitude.

 The  paper is organized as follows. In Sec [\ref{sec:secII}] we start by recalling the basic facts and some kinematical considerations involving the benchmark processes  $W^{\pm}\rightarrow e^{\pm}e^{\pm}\mu^{\mp}\nu$, followed by the subprocesses $W^{\pm}\rightarrow e^{\pm}N$ and $N\rightarrow e^{\pm}\mu^{\mp}\nu$.  We analyze  a specific forward-backward asymmetry for the   lepton number conserving  (LNC) mode $W^{+}\rightarrow e^{+}e^{+}\mu^{-}\nu_e$ and the lepton number violating (LNV) mode $W^{+}\rightarrow e^{+}e^{+}\mu^{-}\overline{\nu}_{\mu}$. This asymmetry  allows to distinguishing between Dirac and Majorana neutrinos in these purely leptonic decays at the LHC. 
  In section [\ref{sec:secIV}] we discuss the distinction of the two same sign charged leptons in the final state at the LHC, which is crucial in order to  extract the asymmetry from the experiment.  
  In Sec [\ref{conclusions}] we conclude with a discussion of our results.

\section{The decay rates and the forward-backward asymmetry}
\label{sec:secII}

In this section we present the theoretical expressions for the purely leptonic processes $W^\pm \to \ell^\pm\ell^\pm\ell^{\prime \mp}\nu$, the forward-backward asymmetry of the 
unlike-charge lepton $\ell^{\prime\mp}$ and the analysis that indicates this asymmetry should be related to the Dirac/Majorana nature of the intermediate neutrino $N$.

 \begin{figure}
     \includegraphics[scale=0.3]{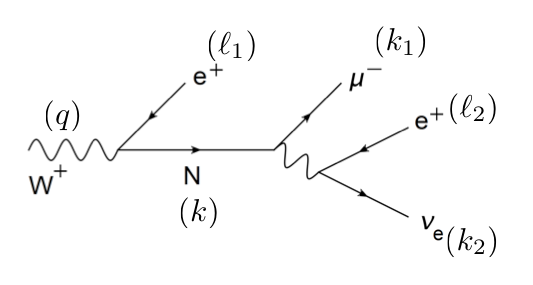} 
     
     \includegraphics[scale=0.3]{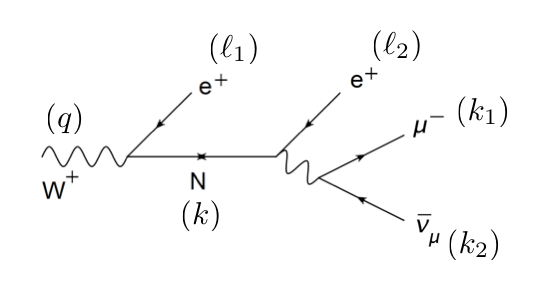} 
  \caption{ \emph{Top:} LNC decay of a $W^+$ mediated by either a Dirac  or Majorana neutrino $N$. \emph{Bottom:} LNV decay of a $W^+$ mediated by a Majorana neutrino only. The momenta are denoted by the symbols in parentheses.}
  \label{fig:DNDecay}
\end{figure}

The part of the lagrangian relevant for this process is given by
\begin{align}
\mathcal{L}=\frac{g}{\sqrt{2}}\sum_{l=e,\mu\tau} W_{\mu}\bar{N}^c U^{\dagger}_{Nl} \gamma^{\mu}P_Ll+ h.c. +...
\end{align}
where $P_L= (1-\gamma_5)/2$, $U_{lN}$ is the mixing matrix of the charged leptons, $l=e,\mu,\tau$ and $N$ denotes the heavy neutrino. 

For simplicity, let us focus on the mode $W^+\to e^+e^+\mu^-\nu$; the  modes with other  flavor and charge are similar. Fig. \ref{fig:DNDecay} shows the two possible processes for this mode, considering that the final neutrino, $\bar\nu_\mu$ or $\nu_e$, is not observable. The upper and lower diagrams correspond to a lepton number conserving (LNC) and a lepton number violating (LNV) process, respectively.

The rates of these two processes, in the notation of Fig.~\ref{fig:DNDecay} and neglecting the masses of the final leptons, can be given in the narrow width approximation for $N$ as:
\begin{eqnarray}
&&\Gamma(W^+\to e^+ e^+ \mu^- \nu_e)_{(LNC)} = \frac{64\sqrt{2}G_F^3}{3}  \frac{M_W^5}{m_N \Gamma_N} 
\nonumber
\\
&&\hspace{12pt} \times    |U_{N e}U_{N \mu}|^2  
   \int d\Phi_{2}  \int d\Phi_{3} \  \frac{(k_1\cdot k_2)}{(M_W^2 - m_N^2 + 2 k\cdot k_1)^2}
\nonumber 
\\  
&&\hspace{12pt}\times 
 \Bigg\{ 2 (k\cdot \ell_2) \Big[ ( k \cdot \ell_1) + \frac{2}{M_W^2}  (q\cdot k )  (q\cdot \ell_1) \Big] 
\nonumber
\\
&&\hspace{36pt} 
 - m_N^2 \Big[   (\ell_2 \cdot \ell_1) + \frac{2}{M_W^2}  (q\cdot \ell_2 )  (q\cdot \ell_1) \Big]
   \Bigg\} 
\label{LNCrate}   
\end{eqnarray}
and
\begin{eqnarray}
&&\Gamma(W^+\to e^+ e^+ \mu^- \bar\nu_\mu)_{(LNV)} =\frac{64\sqrt{2}G_F^3}{3} \frac{M_W^5 m_N} {\Gamma_N}
\nonumber 
\\
&&\hspace{24pt}\times  |U_{N e}|^4 
 \int d\Phi_{2} \int d\Phi_{3}  \ \frac{(k_2 \cdot \ell_2) }{(M_W^2 - m_N^2 - 2 k\cdot \ell_2)^2}
 \nonumber 
 \\  
&&  \hspace{24pt} \times \Big\{ ( k_1 \cdot \ell_1) + \frac{2}{M_W^2}  (q\cdot k_1 )  (q\cdot \ell_1) \Big\} ,
\label{LNVrate}
\end{eqnarray}
where $d\Phi_{2}$ denotes the 2-particle phase space of the first vertex and $d\Phi_{3}$ the 3-particle phase space of the $N$ decay \cite{Patrignani:2016xqp}. 

Since a Dirac $N$ will only produce the LNC process while a Majorana $N$ will produce both LNC and LNV, we can find ways to detect the nature of $N$ by distinguishing between these two processes at the LHC. One cannot distinguish them by the rates because the mixings $|U_{N e}|$ and $|U_{N\mu}|$ are not known \emph{a priori}, nor can we distinguish them by the final particles, because the final $\nu_e$ and $\bar\nu_\mu$ escape detection. However, due to the chiral character of the weak interactions and the fact that $\mu^-$ is in a different fermion line (see Fig.~\ref{fig:DNDecay}), there will be a difference in the angular distribution of the muon momentum.   Indeed, in the $N$ rest frame, if we define our z axis along the initial $W^+$ momentum $\textbf q$, and we call $\theta$ the polar angle of the muon momentum $\textbf k_1$ along this axis, i.e. \hbox{$\cos\theta = \textbf q \cdot \textbf k_1/(|  \textbf q| | \textbf k_1 |)$} the angular distributions of the muon along the $W^+$ direction in the $N$ rest frame are indeed different in the LNC and LNV processes. Calling $x_N \equiv m_N/M_W$, from Eqs.~(\ref{LNCrate}) and (\ref{LNVrate}) we get:

\begin{eqnarray}
&&\frac{d\Gamma}{d\cos\theta} (W^+\to e^+ e^+\mu^-\nu_e)_{(LNC)} =
 {\cal C} \times  |U_{N e}U_{N \mu}|^2 
\nonumber\\
&& \hspace{24pt} \bigg\{ 1 - \cos\theta  \times A\left(x_{N}\right)  \left(  \frac{2- x_N^2}{2+ x_N^2} \right)
\bigg\} 
\label{gammaLNC}
\end{eqnarray}
and
\begin{eqnarray}
&&\frac{d\Gamma}{d\cos\theta} (W^+\to e^+ e^+\mu^-\bar\nu_\mu)_{(LNV)} = {\cal C} \times  |U_{N e}|^4
\nonumber\\
&& \hspace{24pt}  \bigg\{ 1 - \cos\theta \left(  \frac{2- x_N^2}{2+ x_N^2} \right)
\bigg\} ,
\label{gammaLNV}
\end{eqnarray}
%

The global factor ${\cal C}$ is
\begin{align}
{\cal C} \equiv &\frac{\sqrt{2}}{288(2\pi)^4} G_F^3 \frac{M_W^8} { \Gamma_N}  \times F_{W}\left(x_{N}\right)
\label{C2}
\end{align} 
where we defined  the function 
\begin{align}
F_{W}(x_{N}) &\equiv  \frac{(1-x_{N}^2)^2 (2+x_{N}^2)}{x_{N}^3} \bigg\{ \left(6 -3 x_{N}^2- x_{N}^4\right) x_{N}^2 
 \nonumber\\
 &\hspace{24pt}
 + 6 \left(1- x_{N}^2\right) \ln\left(1-x_{N}^2\right)  \bigg\}
 \label{F} ,
\end{align}
and where the neutrino width, $\Gamma_N$, for $m_N \gtrsim 10$ GeV can be estimated as \cite{Atre:2009rg, Dib:2015oka}:
\begin{equation}
\Gamma_{N}\simeq 1.1 \times \frac{G_{F}^{2}}{12\pi^{3}}M_{W}^{5} F_{N}(x_{N})\sum_{\ell} |U_{N \ell}|^{2} ,
\label{eq:GN1}
\end{equation}
where we have included the function $F_N(x_N)$ 
\begin{align}
F_{N}(x_{N})=& \frac{2}{x_{N}^3}\bigg\{ \left(6 -3 x_{N}^2- x_{N}^4\right) x_{N}^2 \nonumber\\
&\hspace{24pt}+ 6 \left(1- x_{N}^2\right) \ln\left(1-x_{N}^2\right) \bigg\} .
\end{align}
This function $F_N(x)$ is due to the momentum dependence of the $W$ propagator, and is an improvement in the $m_N$ 
dependence of $\Gamma_N$ estimate of Refs.~\cite{Atre:2009rg, Dib:2015oka}, where the $W$ propagator was taken as a point interaction. 
In the limit $x_{N}\ll 1$,  $F_W(x_{N}) /F_N(x_N) \to 1$.

\begin{figure}[h!]
\includegraphics[scale=0.6]{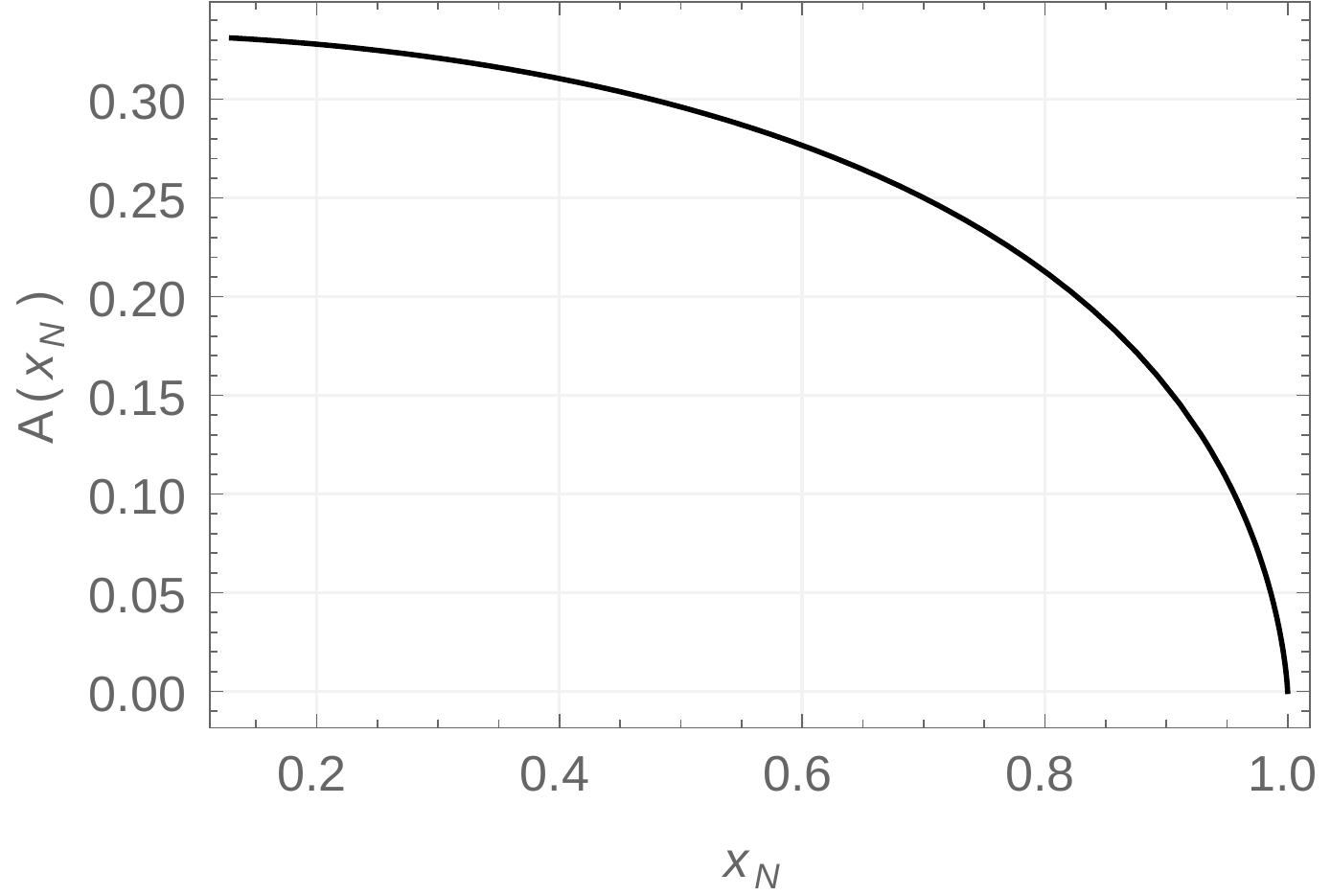} 
\caption{Analyzing power function $A(x_{N})$ given in Eq.~(\ref{AFB}). $x_{N}= m_N/M_W$ }
\label{fig:afb}
\end{figure}

In turn, the  factor $A(x_{N})$ of the angle-dependent term in the LNC distribution is given by: 
\begin{align}
A(x_{N})
&=     \left(x_{N}^2-1 \right)                    
\nonumber \\ 
&\times \frac{ \left(6 - x_{N}^2\right) x_{N}^2+\left(6 - 4 x_{N}^2\right) \ln \left(1-x_{N}^2\right)}{
   \left(6 - 3 x_{N}^2 - x_{N}^4\right) x_{N}^2 +    6 \left(1 -  x_{N}^2\right) \ln \left(1-x_{N}^2\right) }
\label{AFB}
\end{align}
and it is called the \emph{analyzing power}, as it modulates the angular 
dependence of the distribution  \cite{Tait:2000sh,Han:2012vk}.
In the limit  $x_{N}\ll 1$, $A(x_{N}) \to 1/3$, 
while for $x_{N}\rightarrow 1$, $A(x_{N}) \to 0$.
 In Fig.~\ref{fig:afb} we show the behavior of $A(x_{N})$  as a function of $x= m_N/M_W$. 


The difference between the LNC and LNV angular distributions of Eqs.~(\ref{gammaLNC}) and (\ref{gammaLNV})  is essentially the analyzing power factor $A(x_{N})$ in the LNC expression. This difference can be explained as follows. The neutrino $N$ produced in the decay $W^+\to e^+ N$ must be of left handed \emph{chirality} because of the weak current (in the same way, the produced positron in this decay must be right handed). 

\begin{figure}[h!]
\centering
  \includegraphics[scale=0.5]{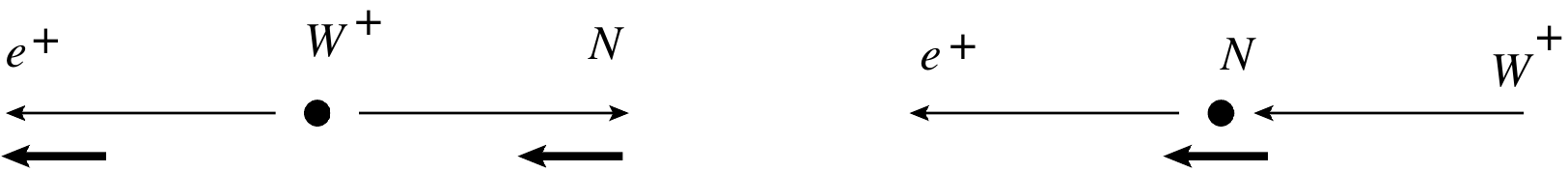}
  \caption{$W^+\to e^+ N$ in two frames: the $W$ rest frame (left); the $N$ rest frame (right). Notice that the $N$ spin direction is conserved under the boost.}
  \label{polariz}
\end{figure}

Provided that $m_N$ is considerably smaller than  $M_W$, the \emph{helicity} of $N$ will be mostly left handed as well (in the $N$ rest frame, the $N$ spin will be polarized along the $W$ momentum $\textbf q$ --see Fig.~\ref{polariz}). For the calculation of the primary process $W^+\to e^+ N$, where $N$ has a polarization vector $s^\mu$ (this vector satisfies $s\cdot s = -1$ and $s\cdot k =0$, where $k$ is the $N$ momentum), one obtains the following squared matrix element:

\begin{align}
|\overline {\cal M}|^2 &= \frac{4 M_W^2 G_F}{3\sqrt{2}} |U_{N\ell}|^2 
\left(  
(k\cdot \ell) + \frac{2}{M_W^2} (q\cdot k) (q\cdot \ell) \right.
\nonumber\\
&\left. - m_N (s\cdot \ell)  - \frac{2 m_N}{M_W^2}(q\cdot s) (q\cdot \ell)
\right) 
\end{align}
Given a neutrino momentum $k=(E,0,0,\textrm{k})$, the right and left helicities of $N$ correspond to $s= \pm (\textrm{k},0,0,E)/m_N$, respectively. It is then easy to show that the  probabilities of the two helicities are in the ratio 
$
|\overline {\cal M}|^2_{\rm right} :    |\overline {\cal M}|^2_{\rm left} =   m_N^2   : 2 M_W^2
$,
and so the probabilities to produce $N$ with right handed or left handed hecility are
\begin{equation}
P_{\rm right} = \frac{m_N^2}{2 M_W^2 + m_N^2}; \quad 
P_{\rm left} = \frac{2 M_W^2}{2 M_W^2 + m_N^2}.
\label{probabilities}
\end{equation}

Now, in the decay of a polarized $N$,  the polarization vector sets a direction along which an anisotropic muon emission occurs (for an unpolarized $N$ the muon emission is isotropic). Moreover, since the muon comes from a different weak current in the LNC and LNV processes (see Fig~\ref{fig:DNDecay}), the anisotropic distributions in these two processes are different. 
We must point out  that the predominant left handed polarization of the produced $N$ occurs regardless of its Dirac or Majorana nature. The latter has an effect in the subsequent decay of $N$. For a Majorana $N$, both decays
\begin{align}
& N \rightarrow \mu^- e^+\nu_e, \label{majo1} \\
& N \rightarrow e^+ \mu^-\bar{\nu}_{\mu}
\label{majo2}
\end{align}
are allowed, while for a Dirac $N$ produced in $W^+\to e^+ N$, the latter mode is forbidden.

Now, these decays, for a polarized $N$  have the following $\theta$ distributions: 
\begin{equation}
\frac{d\Gamma}{d\cos\theta} (N_{pol}\to \mu^- e^+ \nu_e) \sim  |U_{N\mu}|^2 \Big\{ 1 + A\left(x_{N}\right)  \cos\theta \Big\} \label{NLNC}
\end{equation}
and 
\begin{equation}
\frac{d\Gamma}{d\cos\theta} (N_{pol}\to e^+ \mu^-  \bar\nu_\mu) \sim  |U_{N e}|^2 \Big\{ 1 -  \cos\theta \Big\}, \label{NLNV}
\end{equation}

where $\theta$ is, as before, the angle between the momentum of the muon and the polarization vector of $N$ at rest. Here it is clear that the two angular distributions are different precisely because the muon is attached to a different fermion line in the weak interaction. Now, to get to the distributions of the full LNC and LNV processes of Eqs.~(\ref{gammaLNC}) and (\ref{gammaLNV}) we must weigh the polarized $N$ decays by the probabilities in Eq.~(\ref{probabilities}) of the two longitudinal polarizations produced in the primary decay $W\to e^+ N$. The result coincides with  the distributions of Eqs.~(\ref{gammaLNC}) and (\ref{gammaLNV}).  

\bigskip

 Given the fact that the LNC and LNV muon angular distributions are different, we propose to use the following forward-backward asymmetry to try to distinguish between the Dirac vs. Majorana character of the intermediate neutrino $N$ that induces the events:
 
\begin{align}
 A_{FB}= &\frac{N(\cos\theta>0)-N(\cos\theta<0)}{N(\cos\theta>0)+N(\cos\theta<0)}
 \nonumber\\
=& \frac
{\int_0^{1}\frac{d\Gamma}{d\cos\theta}d(\cos\theta)-\int^{0}_{-1}\frac{d\Gamma}{d\cos\theta}d(\cos\theta)}
{\int_{-1}^{1}\frac{d\Gamma}{d\cos\theta}d(\cos\theta)} ,
\label{AFB0}
 \end{align}
where $N(\cos\theta > 0)$ and $N(\cos\theta < 0)$ denote the number events with the muon moving forwards or backwards with respect to the decaying $W$, in the $N$ rest frame of the heavy neutrino. 

If $N$ is a Dirac neutrino, only the LNC process occurs, in which case the  analyzing power $A(m_N/M_W)$ 
determines directly the outcome of the experimental forward-backward asymmetry, regardless of lepton mixing
(although the mixing determines the total number of events): 

 \begin{align}
 A_{FB}^{(Dirac)} =-\frac{1}{2}A\left(x_{N}\right) \left(  \frac{2- x_N^2}{2+ x_N^2} \right).  
 \label{AFB_Dirac}
 \end{align}

Instead, if $N$ is a Majorana neutrino, the rate of events $W\to e^+e^+\mu^-\nu$ is the sum of the LNC and LNV rates given in Eqs.~(\ref{gammaLNC}) and (\ref{gammaLNV}), and then the forward-backward asymmetry will depend not only on the analyzing power of the LNC component, but also on the relative lepton mixing elements $|U_{Ne}|$ and $|U_{N\mu}|$: 

\begin{align}
A_{FB}^{(Maj)}
=&  - 
  \frac{A\left(x_{N}\right)   |U_{N \mu}|^2  +  |U_{N e}|^2  }{2\Big( |U_{N \mu}|^2+|U_{N e}|^2 \Big)}
  \left(  \frac{2- x_N^2}{2+ x_N^2} \right) .
\label{AFB_Maj}
\end{align}

The Dirac asymmetry, Eq.~(\ref{AFB_Dirac}), is independent of the mixings and,  due to the function $A(x_{N})$ (see Fig. \ref{fig:afb}),  is approximatively  $-1/6$ for a wide range of $m_N$, except if $m_N$ approaches $M_W$, in which case it tends to vanish. 

In contrast, the Majorana asymmetry, Eq.~(\ref{AFB_Maj}),  does depend on the mixings and in general it is larger than in the Dirac case. For events $e^+e^+\mu^-$, if $|U_{N\mu}| \gg |U_{Ne}||$,  it tends to the same value as in the Dirac case, because it is dominated by the LNC contribution, but in the opposite limit, namely  $|U_{N\mu}| \ll |U_{Ne}|$, dominated by the LNV contribution, it is larger: close to $-1/2$ for most values of $m_N$, decreasing to $-1/6$ as $m_N$ approaches $M_W$. 

Consequently one can in principle distinguish the Dirac from the Majorana case using this asymmetry. However, notice
that in the case $|U_{N\mu}| \gg |U_{Ne}|$, for events $e^\pm e^\pm\mu^\mp$ the asymmetry is the same for both Dirac and Majorana $N$, because the Majorana asymmetry is dominated by the LNC contribution. 

In this case the channels $\mu^\pm\mu^\pm e^\mp$ are more appropriate for discrimination, because now the Majorana asymmetry is dominated by the LNV contribution. The angular distribution in this case is that of the electron flavor instead of the muon.

  \begin{figure*}[t]
  \includegraphics[scale=0.5]{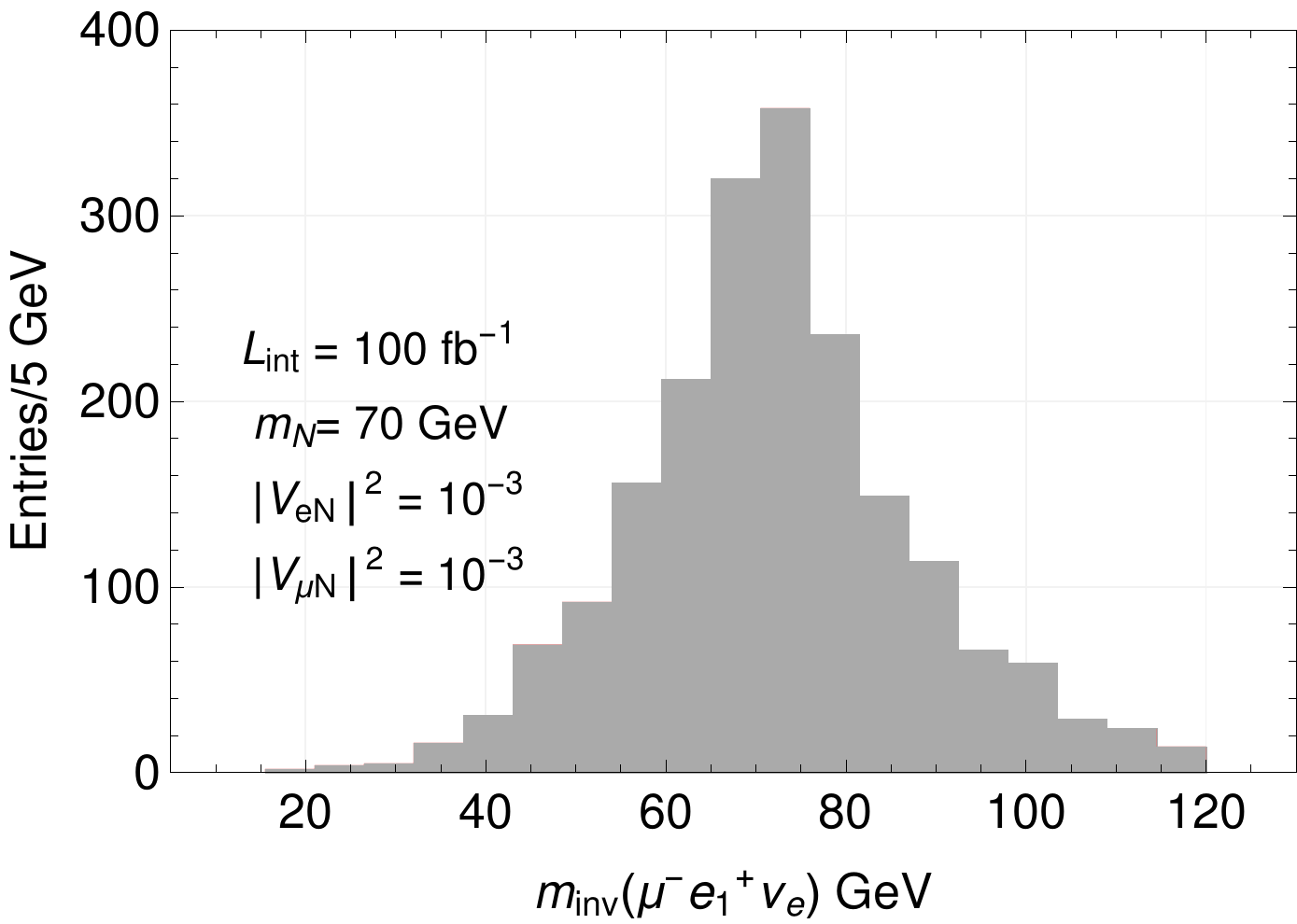} 
  \includegraphics[scale=0.5]{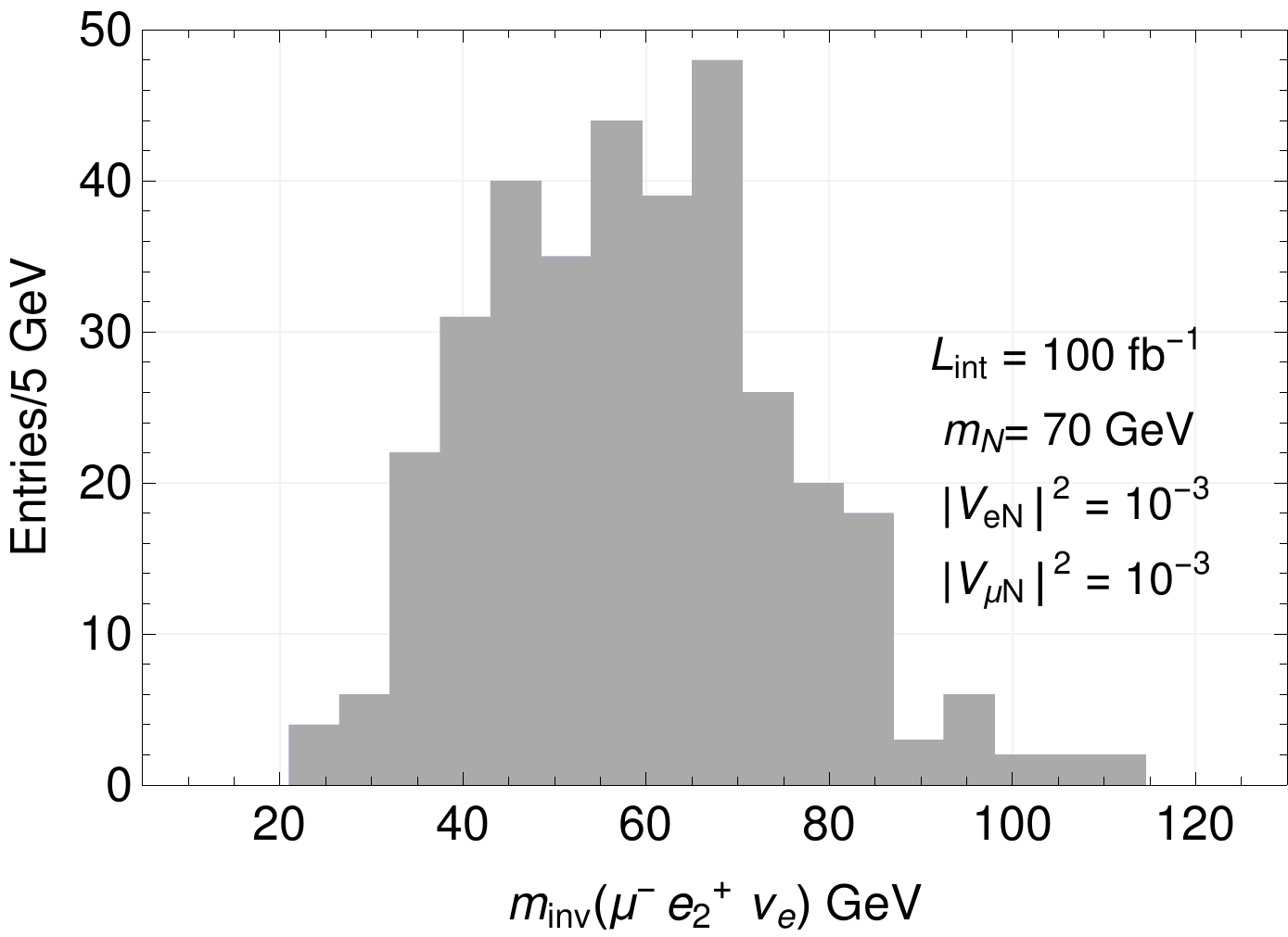}\\
    \includegraphics[scale=0.5]{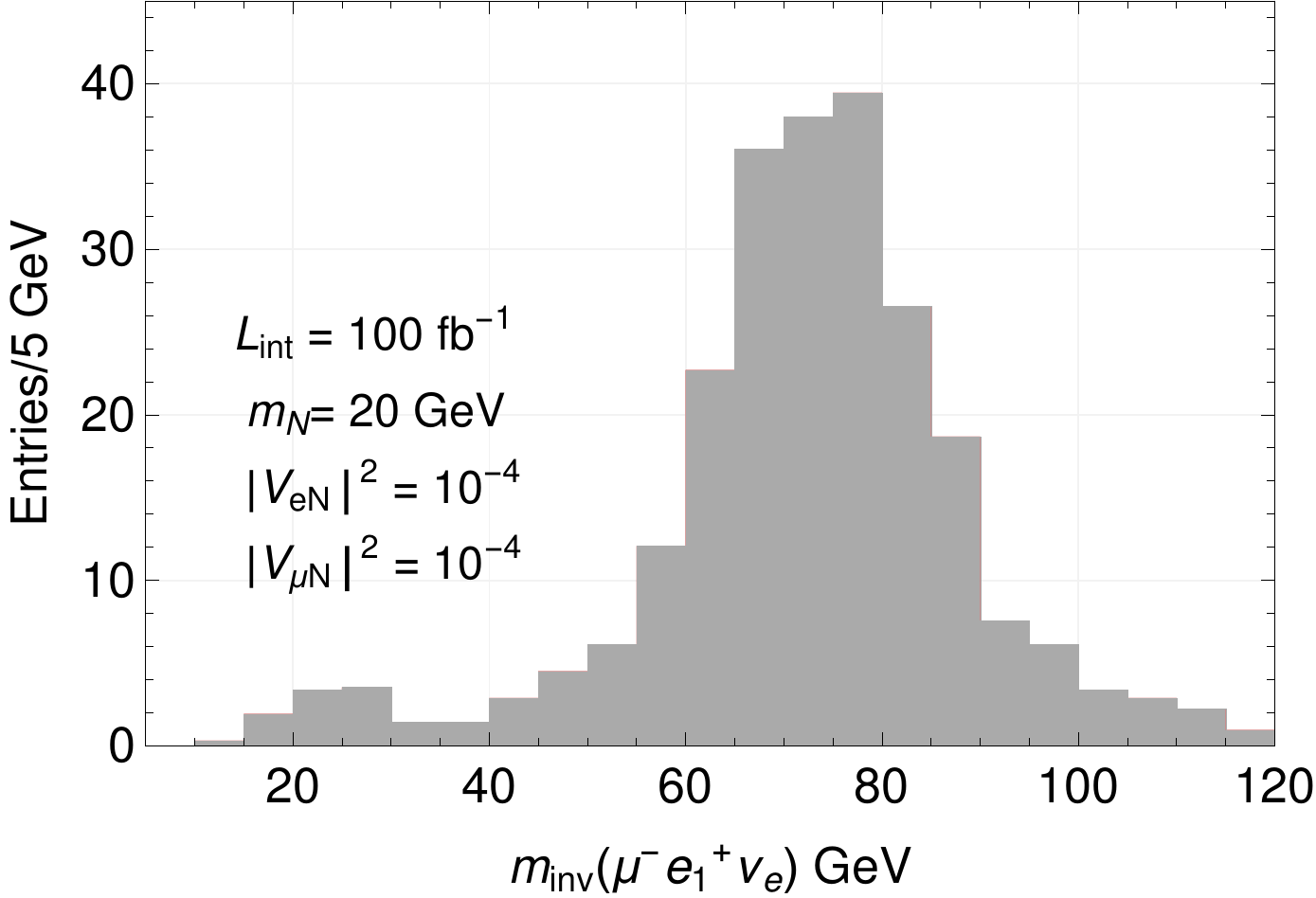} 
  \includegraphics[scale=0.5]{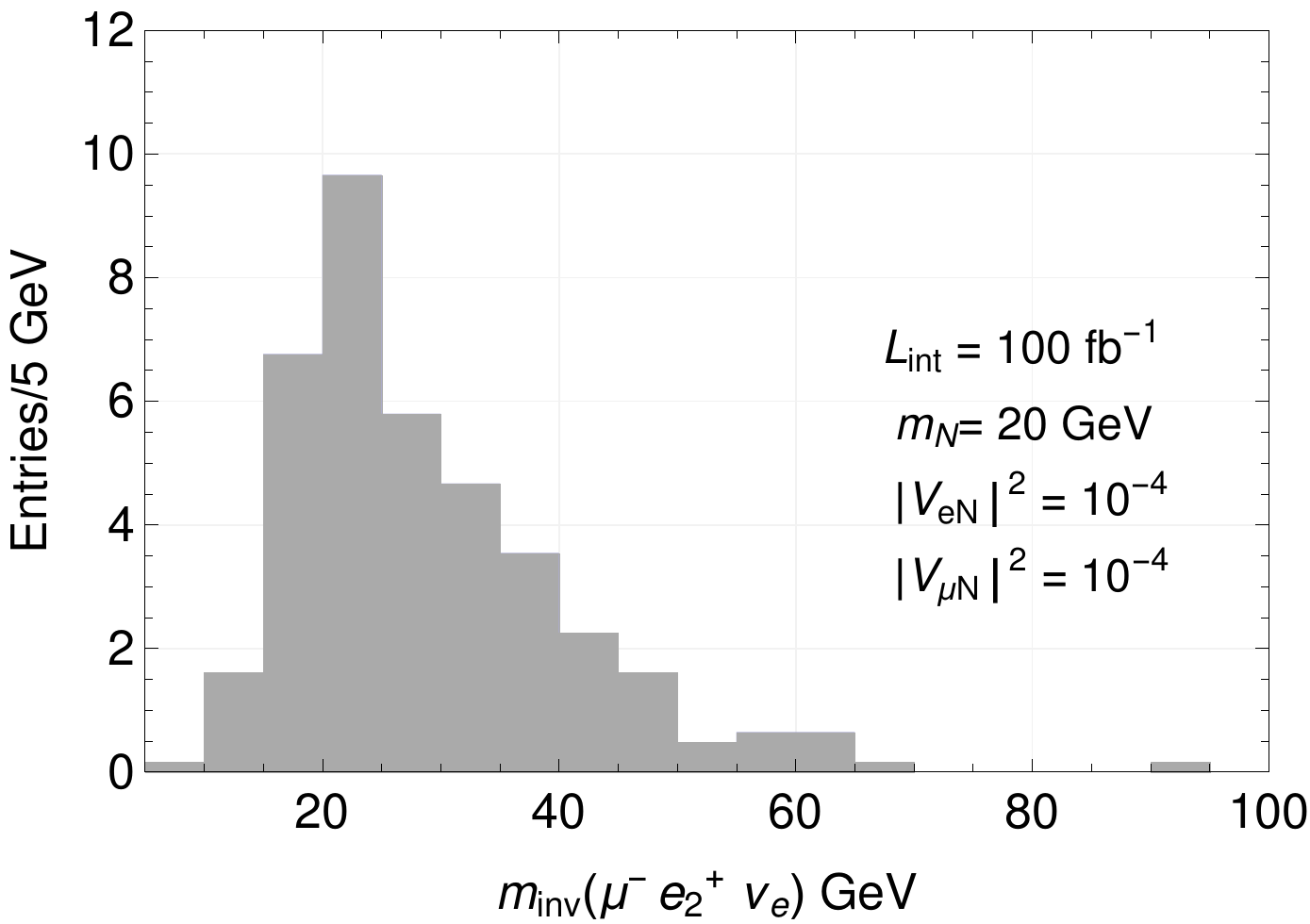} 
  \caption{Invariant mass distributions for the  $e^+_1\mu^-\nu_e$ and $e^+_2\mu^-\nu_e$ invariant masses and for several heavy neutrino masses with center of mass energy of $\sqrt{s}=13$ TeV. }
  \label{fig_e1e2}
\end{figure*}

\section{Distinguishing the same-sign leptons at the LHC} \label{sec:secIV}

In order to experimentally  determine the asymmetries of Eqs. \eqref{AFB_Dirac} and \eqref{AFB_Maj}, one needs to identify which of the two equal-sign leptons originates from the first vertex and from the second vertex (see  Fig. \ref{fig:DNDecay}). From this identification one can do the appropriate boosts to the $W$ rest frame or to the $N$ rest frame, as required. In this section we study this issue of telling apart the two equal-sign leptons (see also Ref.~\cite{Dib:2016wge}).

As an example, consider the LNC production and decay channel $p p \rightarrow W^+ \rightarrow e^+ N \rightarrow e^+ \mu^-e^+\nu_e$, in which two positrons are in the final state. In the LNV channel there is one $\bar{\nu}_{\mu}$ in the final state, but as we shall see, the considerations we use apply to the latter case as well.

In Fig.~\ref{fig_e1e2} we show the invariant masses for the $e^+_1\mu^-\nu_e$ and $e^+_2\mu^-\nu_e$, where $e_1^+$ and $e_2^+$ are the harder and softer positron sorted by  energy, respectively. The events were  generated at LO using MadGraph 5~\cite{Alwall:2014hca}, hadronized with Pythia 6~\cite{Sjostrand:2006za} and passed to Delphes 3~\cite{deFavereau:2013fsa} for detector simulation. In the figures we may see that when the heavy neutrino mass is above 70 GeV the hardest positron reconstructs better its mass. This is in agreement with the expectation that in this case the heavy neutrino carries most of the energy.

Conversely, when the heavy neutrino mass is below 40 GeV, it is the softer positron that reconstructs the  heavy neutrino mass -- as can be seen in Fig. \ref{fig_e1e2}. Once again, this is in agreement with the expectation that  the heavy neutrino carries less energy than in the more massive case, since in the massless limit is would have exactly half of the energy taken from the W decay. We find that in the  region between 50-70 GeV both electrons show similar behavior in the $m_N$ reconstruction and hence cannot be distinguished with the above strategy. There is another interesting region, which is when $m_N \leq 10$ GeV. In this region the decay length of the heavy neutrino becomes visible in the form of a displaced vertex inside the detector. Clearly in this case the distinction of both positrons will be obvious.  The background processes and the possibility of distinguishing both same-sign leptons have been extensively studied in \cite{Dib:2016wge,Dib:2017iva,Dib:2017vux}. Therefore we do not discuss this issue further here. In summary, for
 $m_N\lesssim 10$~GeV, the two leptons can be separated due to vertex displacement. For higher masses no vertex displacement is observable since $N$ is too short-living. In those cases the separation of the two leptons is less certain. As shown, statistically the primary lepton tends to be the most energetic one if $m_N\lesssim 50$ GeV and the less energetic one if $m_N \gtrsim 70$ GeV. In the mass region between 50 and 70 GeV no such distinction occurs.
 
 Another important experimental issue here is the expected number of events. For a realistic analysis that involves angular distributions, one would need a rather large sample of these rare events. According to Ref.~ \cite{Aad:2016naf}, at the end of the LHC Run II, with an integrated luminosity of 120~fb$^{-1}$, ATLAS is expected to have of the order of $2\times 10^8$ leptonic $W$ decays into $e \nu$ and $\mu \nu$. Considering that the $W$ width is $\Gamma_W \sim 2$~GeV and the branching ratio into each lepton is near 10\%,  using Eqs.~(\ref{gammaLNC}--\ref{C2}) and (\ref{eq:GN1}), we can estimate the expected number of $W\to e^+ e^+ \mu^-\nu$  or 
 $W\to \mu^+ \mu^+ e^-\nu$ events. Without considering the impact of systematic errors or backgrounds, the expected numbers of events for the Majorana and Dirac scenarios are estimated to be:
\begin{align}
N(W\rightarrow e^{+}e^{+}\mu^{-}\nu)_{Maj}&\sim 6 \times 10^{6} 
\\
&\hspace{-36pt} \times (|U_{Ne}|^2+|U_{N \mu}|^2)|U_{Ne}|^2  /{ \sum_{l}|U_{Nl}|^2}  
\nonumber\\ 
N(W\rightarrow e^{+}e^{+}\mu^{-}\nu)_{Dirac}& \sim 6 \times 10^{6}
\\
&\hspace{-36pt}\times  |U_{Ne} U_{N\mu}|^2/ {\sum_{l}|U_{Nl}|^2}
\nonumber
\end{align}

From these expressions one can deduce that the largest of the mixings $ |U_{N e}|^2$ and $ |U_{N \mu}|^2$ must be 
at least of order $10^{-6}$ for the LHC Run II to get a positive signal of these modes.  Take for example the case  $|U_{Ne}| \gg |U_{N \mu}| \sim |U_{N\tau}|$; then  the above reduces to:
\begin{align}
&N(W\rightarrow e^{+}e^{+}\mu^{-}\nu)_{Maj} \sim 6 \times 10^{6} |U_{N e}|^2
\nonumber
\\ 
&N(W\rightarrow e^{+}e^{+}\mu^{-}\nu)_{Dirac} \sim 6 \times 10^{6} |U_{N \mu}|^2 ,
\nonumber
\end{align}
 implying that the expected number of these events for  $|U_{Ne}|^2 \sim 10^{-6}$ is 6 (0) for Majorana (Dirac) $N$.  
 
 Alternatively, if $|U_{Ne}| \sim |U_{N \mu}| \sim |U_{N \tau}|$, the above reduces to:
\begin{align}
&N(W\rightarrow e^{+}e^{+}\mu^{-}\nu)_{Maj} \sim 4 \times 10^{6} |U_{Ne}|^2 \nonumber \\ \nonumber
&N(W\rightarrow e^{+}e^{+}\mu^{-}\nu)_{Dirac} \sim 2 \times 10^{6}  |U_{Ne}|^2
\end{align}
and the expected number of events for  $|U_{Ne}|^2 \sim 10^{-6}$ would be around 4 (2) for Majorana (Dirac) $N$. 

Finally, if $|U_{N\mu}| \gg |U_{N e}|\sim |U_{N \tau}|$, the number of $e^\pm e^\pm \mu^\mp$ events for both Dirac and Majorana cases are much smaller, in which case the $\mu^{\pm} \mu^{\pm} e^{\mp}$ events should be considered.
Therefore, for an integrated luminosity of 120 fb$^{-1}$ at the LHC Run II,  $W\rightarrow e^{\pm}e^{\pm}\mu^{\mp}$ events can be found provided $|U_{Ne}|^2 \gtrsim 10^{-6}$ while $W\rightarrow \mu^{\pm}\mu^{\pm}e^{\mp}$ signals can be found if $|U_{N\mu}|^2 \gtrsim  10^{-6}$. These are the limits in cases where $m_N < M_W$, but as $m_N$ approaches $M_W$ these theoretical rates are smaller and the lowest $|U_{N\ell}|^2$ for detection become larger. 
Moreover, in cases where $|U_{N e}|^2 \sim |U_{N\mu}|^2$, the discrimination between Dirac and Majorana $N$ requires the forward-backward asymmetry of Eq.~\eqref{AFB0}. In order to build this asymmetry one needs more data than the minima deduced above, hence $|U_{N\ell}|^2$ must be above $10^{-6}$ or the $W\to \ell\nu$ data sample must be larger than what is expected at the end of the LHC Run II. On the other hand, current upper bounds for $|U_{N e}|^2$ and $|U_{N\mu}|^2$ for $N$ with mass in the range 10~GeV~$< m_N< M_W$ are near $10^{-5}$ \cite{Abreu:1996pa}, leaving about an order of magnitude in the range of the mixings to explore in the  LHC Run II.

\noindent

\section{Conclusions}
\label{conclusions}
We presented a new method to discriminate Dirac vs Majorana sterile neutrinos with masses $m_{N}\leq m_{W}$ that would induce $W$ trilepton decays $e^\pm e^\pm\mu^\mp$ and $\mu^\pm\mu^\pm e^\mp$ at the LHC. A simple strategy based on a forward-asymmetry is proposed. This asymmetry is different depending on the Dirac/Majorana character of the heavy neutrino.  If the heavy neutrino is a Dirac particle, the forward-backward asymmetry depends on the heavy neutrino mass but not on the heavy-to-light mixing parameters. Instead, for a Majorana neutrino the asymmetry depends on the both the mass and the mixing matrix elements. The method presented here is complementary to the strategy proposed in \cite{Dib:2016wge,Dib:2017iva, Dib:2017vux}, in the sense that it  allows to distinguish the neutrino nature also in the case where the neutrino-lepton mixing elements are of the same order. 

In order to construct the asymmetry, a crucial point is to identify the primary and secondary leptons of same charge in the final state $\ell^\pm\ell^\pm\ell^{\prime \mp}$. 
 Using invariant mass analysis distributions we find ranges for $m_{N}$ where the LHC could distinguish among these two leptons. 

 If $m_{N} \lesssim 50$ GeV ($m_{N} \gtrsim 70$ GeV) the first positron tends to be the most (least) energetic, while in the region $50 \lesssim m_{N} \lesssim 70$GeV, the order of the two same-sign leptons can not be distinguished.

We estimated the number of these events for the integrated luminosity of 120 fb$^{-1}$ expected at the end of the LHC Run II, and found that both Majorana and Dirac signals could be found provided that at least one the mixings $|U_{N \mu}|^2$ and $|U_{N e}|^2$ is $10^{-6}$ or above.

\label{sec:concl}
\section*{Acknowledgments}

J.V. thanks Goran Senjanovi\'c for useful and enlightening discussions. 
This work is supported in part by Fondecyt (Chile) grants No. 3170154, No  1180232 and No. 1170171, and by CONICYT (Chile) Ring ACT1406 and PIA/Basal FB0821.

------------------------------------------------------


%
\bibliographystyle{jhep}
\bibliography{biblio}

\providecommand{\href}[2]{#2}\begingroup\raggedright\begin{thebibliography}{10}

\bibitem{PhysRevLett.81.1562}
{\scshape Super-Kamiokande Collaboration} collaboration, Y.~Fukuda,
  T.~Hayakawa, E.~Ichihara, K.~Inoue, K.~Ishihara, H.~Ishino et~al.,
  \emph{Evidence for oscillation of atmospheric neutrinos},
  \href{http://dx.doi.org/10.1103/PhysRevLett.81.1562}{\emph{Phys. Rev. Lett.}
  {\bfseries 81} (Aug, 1998) 1562--1567}.

\bibitem{Ahmad:2002jz}
{\scshape SNO} collaboration, Q.~R. Ahmad et~al., \emph{{Direct evidence for
  neutrino flavor transformation from neutral current interactions in the
  Sudbury Neutrino Observatory}},
  \href{http://dx.doi.org/10.1103/PhysRevLett.89.011301}{\emph{Phys. Rev.
  Lett.} {\bfseries 89} (2002) 011301},
  [\href{https://arxiv.org/abs/nucl-ex/0204008}{{\ttfamily nucl-ex/0204008}}].

\bibitem{Eguchi:2002dm}
{\scshape KamLAND} collaboration, K.~Eguchi et~al., \emph{{First results from
  KamLAND: Evidence for reactor anti-neutrino disappearance}},
  \href{http://dx.doi.org/10.1103/PhysRevLett.90.021802}{\emph{Phys. Rev.
  Lett.} {\bfseries 90} (2003) 021802},
  [\href{https://arxiv.org/abs/hep-ex/0212021}{{\ttfamily hep-ex/0212021}}].

\bibitem{Forero:2014bxa}
D.~V. Forero, M.~Tortola and J.~W.~F. Valle, \emph{{Neutrino oscillations
  refitted}}, \href{http://dx.doi.org/10.1103/PhysRevD.90.093006}{\emph{Phys.
  Rev.} {\bfseries D90} (2014) 093006},
  [\href{https://arxiv.org/abs/1405.7540}{{\ttfamily 1405.7540}}].

\bibitem{Majorana2008}
E.~Majorana, \emph{Teoria simmetrica dell'elettrone e del positrone},
  \href{http://dx.doi.org/10.1007/BF02961314}{\emph{Il Nuovo Cimento
  (1924-1942)} {\bfseries 14} (2008) 171}.

\bibitem{MINKOWSKI1977421}
P.~Minkowski, \emph{$\mu\rightarrow e \gamma$ at a rate of one out of 109 muon
  decays?},
  \href{http://dx.doi.org/http://dx.doi.org/10.1016/0370-2693(77)90435-X}{\emph{Physics
  Letters B} {\bfseries 67} (1977) 421 -- 428}.

\bibitem{Mohapatra:1979ia}
R.~N. Mohapatra and G.~Senjanovi\'c, \emph{{Neutrino Mass and Spontaneous
  Parity Violation}},
  \href{http://dx.doi.org/10.1103/PhysRevLett.44.912}{\emph{Phys. Rev. Lett.}
  {\bfseries 44} (1980) 912}.

\bibitem{Glashow:1979nm}
S.~L. Glashow, \emph{{The Future of Elementary Particle Physics}},
  \href{http://dx.doi.org/10.1007/978-1-4684-7197-7_15}{\emph{NATO Sci. Ser. B}
  {\bfseries 61} (1980) 687}.

\bibitem{GellMann:1980vs}
M.~Gell-Mann, P.~Ramond and R.~Slansky, \emph{{Complex Spinors and Unified
  Theories}}, {\emph{Conf. Proc.} {\bfseries C790927} (1979) 315--321},
  [\href{https://arxiv.org/abs/1306.4669}{{\ttfamily 1306.4669}}].

\bibitem{Yanagida:1979as}
T.~Yanagida, \emph{{HORIZONTAL SYMMETRY AND MASSES OF NEUTRINOS}}, {\emph{Conf.
  Proc.} {\bfseries C7902131} (1979) 95--99}.

\bibitem{Keung:1983uu}
W.-Y. Keung and G.~Senjanovi\'c, \emph{{Majorana Neutrinos and the Production
  of the Right-handed Charged Gauge Boson}},
  \href{http://dx.doi.org/10.1103/PhysRevLett.50.1427}{\emph{Phys. Rev. Lett.}
  {\bfseries 50} (1983) 1427}.

\bibitem{Ferrari:2000sp}
A.~Ferrari, J.~Collot, M.-L. Andrieux, B.~Belhorma, P.~de~Saintignon, J.-Y.
  Hostachy et~al., \emph{{Sensitivity study for new gauge bosons and
  right-handed Majorana neutrinos in $p p$ collisions at $s$ = 14-TeV}},
  \href{http://dx.doi.org/10.1103/PhysRevD.62.013001}{\emph{Phys. Rev.}
  {\bfseries D62} (2000) 013001}.

\bibitem{Kovalenko:2009td}
S.~Kovalenko, Z.~Lu and I.~Schmidt, \emph{{Lepton Number Violating Processes
  Mediated by Majorana Neutrinos at Hadron Colliders}},
  \href{http://dx.doi.org/10.1103/PhysRevD.80.073014}{\emph{Phys. Rev.}
  {\bfseries D80} (2009) 073014},
  [\href{https://arxiv.org/abs/0907.2533}{{\ttfamily 0907.2533}}].

\bibitem{Han:2012vk}
T.~Han, I.~Lewis, R.~Ruiz and Z.-g. Si, \emph{{Lepton Number Violation and
  $W^\prime$ Chiral Couplings at the LHC}},
  \href{http://dx.doi.org/10.1103/PhysRevD.87.035011,
  10.1103/PhysRevD.87.039906}{\emph{Phys. Rev.} {\bfseries D87} (2013) 035011},
  [\href{https://arxiv.org/abs/1211.6447}{{\ttfamily 1211.6447}}].

\bibitem{Das:2015toa}
A.~Das and N.~Okada, \emph{{Improved bounds on the heavy neutrino productions
  at the LHC}}, \href{http://dx.doi.org/10.1103/PhysRevD.93.033003}{\emph{Phys.
  Rev.} {\bfseries D93} (2016) 033003},
  [\href{https://arxiv.org/abs/1510.04790}{{\ttfamily 1510.04790}}].

\bibitem{Das:2016hof}
A.~Das, P.~Konar and S.~Majhi, \emph{{Production of Heavy neutrino in
  next-to-leading order QCD at the LHC and beyond}},
  \href{http://dx.doi.org/10.1007/JHEP06(2016)019}{\emph{JHEP} {\bfseries 06}
  (2016) 019}, [\href{https://arxiv.org/abs/1604.00608}{{\ttfamily
  1604.00608}}].

\bibitem{Das:2017gke}
A.~Das, P.~Konar and A.~Thalapillil, \emph{{Jet substructure shedding light on
  heavy Majorana neutrinos at the LHC}},
  \href{https://arxiv.org/abs/1709.09712}{{\ttfamily 1709.09712}}.

\bibitem{Das:2017pvt}
A.~Das, \emph{{Pair production of heavy neutrinos in next-to-leading order QCD
  at the hadron colliders in the inverse seesaw framework}},
  \href{https://arxiv.org/abs/1701.04946}{{\ttfamily 1701.04946}}.

\bibitem{Das:2017zjc}
A.~Das, P.~S.~B. Dev and C.~S. Kim, \emph{{Constraining Sterile Neutrinos from
  Precision Higgs Data}},
  \href{http://dx.doi.org/10.1103/PhysRevD.95.115013}{\emph{Phys. Rev.}
  {\bfseries D95} (2017) 115013},
  [\href{https://arxiv.org/abs/1704.00880}{{\ttfamily 1704.00880}}].

\bibitem{Das:2017hmg}
A.~Das, P.~S.~B. Dev and R.~N. Mohapatra, \emph{{Same Sign vs Opposite Sign
  Dileptons as a Probe of Low Scale Seesaw Mechanisms}},
  \href{https://arxiv.org/abs/1709.06553}{{\ttfamily 1709.06553}}.

\bibitem{Aad:2015xaa}
{\scshape ATLAS} collaboration, G.~Aad et~al., \emph{{Search for heavy Majorana
  neutrinos with the ATLAS detector in pp collisions at $ \sqrt{s}=8 $ TeV}},
  \href{http://dx.doi.org/10.1007/JHEP07(2015)162}{\emph{JHEP} {\bfseries 07}
  (2015) 162}, [\href{https://arxiv.org/abs/1506.06020}{{\ttfamily
  1506.06020}}].

\bibitem{Khachatryan:2015gha}
{\scshape CMS} collaboration, V.~Khachatryan et~al., \emph{{Search for heavy
  Majorana neutrinos in $\mu^\pm \mu^\pm+$ jets events in proton-proton
  collisions at $\sqrt{s}$ = 8 TeV}},
  \href{http://dx.doi.org/10.1016/j.physletb.2015.06.070}{\emph{Phys. Lett.}
  {\bfseries B748} (2015) 144--166},
  [\href{https://arxiv.org/abs/1501.05566}{{\ttfamily 1501.05566}}].

\bibitem{Atre:2009rg}
A.~Atre, T.~Han, S.~Pascoli and B.~Zhang, \emph{{The Search for Heavy Majorana
  Neutrinos}},
  \href{http://dx.doi.org/10.1088/1126-6708/2009/05/030}{\emph{JHEP} {\bfseries
  05} (2009) 030}, [\href{https://arxiv.org/abs/0901.3589}{{\ttfamily
  0901.3589}}].

\bibitem{Helo:2013esa}
J.~C. Helo, M.~Hirsch and S.~Kovalenko, \emph{{Heavy neutrino searches at the
  LHC with displaced vertices}},
  \href{http://dx.doi.org/10.1103/PhysRevD.89.073005,
  10.1103/PhysRevD.93.099902}{\emph{Phys. Rev.} {\bfseries D89} (2014) 073005},
  [\href{https://arxiv.org/abs/1312.2900}{{\ttfamily 1312.2900}}].

\bibitem{Izaguirre:2015pga}
E.~Izaguirre and B.~Shuve, \emph{{Multilepton and Lepton Jet Probes of
  Sub-Weak-Scale Right-Handed Neutrinos}},
  \href{http://dx.doi.org/10.1103/PhysRevD.91.093010}{\emph{Phys. Rev.}
  {\bfseries D91} (2015) 093010},
  [\href{https://arxiv.org/abs/1504.02470}{{\ttfamily 1504.02470}}].

\bibitem{Dib:2015oka}
C.~O. Dib and C.~S. Kim, \emph{{Discovering sterile Neutrinos ligther than
  $M_W$ at the LHC}},
  \href{http://dx.doi.org/10.1103/PhysRevD.92.093009}{\emph{Phys. Rev.}
  {\bfseries D92} (2015) 093009},
  [\href{https://arxiv.org/abs/1509.05981}{{\ttfamily 1509.05981}}].

\bibitem{Dib:2016wge}
C.~O. Dib, C.~S. Kim, K.~Wang and J.~Zhang, \emph{{Distinguishing
  Dirac/Majorana Sterile Neutrinos at the LHC}},
  \href{http://dx.doi.org/10.1103/PhysRevD.94.013005}{\emph{Phys. Rev.}
  {\bfseries D94} (2016) 013005},
  [\href{https://arxiv.org/abs/1605.01123}{{\ttfamily 1605.01123}}].

\bibitem{Dib:2017iva}
C.~O. Dib, C.~S. Kim and K.~Wang, \emph{{Signatures of Dirac and Majorana
  sterile neutrinos in trilepton events at the LHC}},
  \href{http://dx.doi.org/10.1103/PhysRevD.95.115020}{\emph{Phys. Rev.}
  {\bfseries D95} (2017) 115020},
  [\href{https://arxiv.org/abs/1703.01934}{{\ttfamily 1703.01934}}].

\bibitem{Dib:2017vux}
C.~O. Dib, C.~S. Kim and K.~Wang, \emph{{Search for Heavy Sterile Neutrinos in
  Trileptons at the LHC}},
  \href{http://dx.doi.org/10.1088/1674-1137/41/10/103103}{\emph{Chin. Phys.}
  {\bfseries C41} (2017) 103103},
  [\href{https://arxiv.org/abs/1703.01936}{{\ttfamily 1703.01936}}].

\bibitem{Tait:2000sh}
T.~M.~P. Tait and C.~P. Yuan, \emph{{Single top quark production as a window to
  physics beyond the standard model}},
  \href{http://dx.doi.org/10.1103/PhysRevD.63.014018}{\emph{Phys. Rev.}
  {\bfseries D63} (2000) 014018},
  [\href{https://arxiv.org/abs/hep-ph/0007298}{{\ttfamily hep-ph/0007298}}].

\bibitem{Patrignani:2016xqp}
{\scshape Particle Data Group} collaboration, C.~Patrignani et~al.,
  \emph{{Review of Particle Physics}},
  \href{http://dx.doi.org/10.1088/1674-1137/40/10/100001}{\emph{Chin. Phys.}
  {\bfseries C40} (2016) 100001}.

\bibitem{Alwall:2014hca}
J.~Alwall, R.~Frederix, S.~Frixione, V.~Hirschi, F.~Maltoni, O.~Mattelaer
  et~al., \emph{{The automated computation of tree-level and next-to-leading
  order differential cross sections, and their matching to parton shower
  simulations}}, \href{http://dx.doi.org/10.1007/JHEP07(2014)079}{\emph{JHEP}
  {\bfseries 07} (2014) 079},
  [\href{https://arxiv.org/abs/1405.0301}{{\ttfamily 1405.0301}}].

\bibitem{Sjostrand:2006za}
T.~Sjostrand, S.~Mrenna and P.~Z. Skands, \emph{{PYTHIA 6.4 Physics and
  Manual}}, \href{http://dx.doi.org/10.1088/1126-6708/2006/05/026}{\emph{JHEP}
  {\bfseries 05} (2006) 026},
  [\href{https://arxiv.org/abs/hep-ph/0603175}{{\ttfamily hep-ph/0603175}}].

\bibitem{deFavereau:2013fsa}
{\scshape DELPHES 3} collaboration, J.~de~Favereau, C.~Delaere, P.~Demin,
  A.~Giammanco, V.~Lemaître, A.~Mertens et~al., \emph{{DELPHES 3, A modular
  framework for fast simulation of a generic collider experiment}},
  \href{http://dx.doi.org/10.1007/JHEP02(2014)057}{\emph{JHEP} {\bfseries 02}
  (2014) 057}, [\href{https://arxiv.org/abs/1307.6346}{{\ttfamily 1307.6346}}].

\bibitem{Aad:2016naf}
{\scshape ATLAS} collaboration, G.~Aad et~al., \emph{{Measurement of $W^{\pm}$
  and $Z$-boson production cross sections in $pp$ collisions at $\sqrt{s}=13$
  TeV with the ATLAS detector}},
  \href{http://dx.doi.org/10.1016/j.physletb.2016.06.023}{\emph{Phys. Lett.}
  {\bfseries B759} (2016) 601--621},
  [\href{https://arxiv.org/abs/1603.09222}{{\ttfamily 1603.09222}}].

\bibitem{Abreu:1996pa}
{\scshape DELPHI} collaboration, P.~Abreu et~al., \emph{{Search for neutral
  heavy leptons produced in Z decays}},
  \href{http://dx.doi.org/10.1007/s002880050370}{\emph{Z. Phys.} {\bfseries
  C74} (1997) 57--71}.

\end{thebibliography}\endgroup

\end{document}